\documentstyle[prl,eqsecnum,aps]{revtex}
\begin{document}
\author{J. Q. Shen\footnote{Author's E-mail address: jqshen@coer.zju.edu.cn}}
\address{Zhejiang Institute of Modern Physics and Department of
Physics, Zhejiang University, Hangzhou 310027, P.R. China}
\date{\today }
\title{A note: graviton spin, gravitomagnetic fields and self-interaction
\\of non-inertial frame of reference\footnote{It will be submitted nowhere else for publication,
just uploaded at the e-print archives. It is a supplement to a brief report (submitted to PRD)
entitled ``The purely gravitational generalization of spin-rotation couplings'' (by J.Q. Shen).}}
\maketitle

\begin{abstract}
Three weak gravitational effects associated with the
gravitomagnetic fields are taken into account in this paper: (i)
we discuss the background Lorentz transformation and gauge
transformation in a linearized gravity theory, and obtain the
expression for the spin of gravitational field by using the
canonical procedure and Noether theorem; (ii) we point out that by
using the coordinate transformation from the fixed
frame\footnote{Note that here the fixed frame of reference is also
viewed as a rest frame. The rotating frame rotates at an angular
velocity, $\vec{\omega}$, relative to the rest frame.} to the
rotating frame, it is found that the nature of Mashhoon's
spin-rotation coupling is in fact an interaction between the
gravitomagnetic moment of a spinning particle and the
gravitomagnetic fields. The fact that the rotational angular
velocity of a rotating frame can be viewed as a gravitomagnetic
field is demonstrated; (iii) a purely gravitational generalization
of Mashhoon's spin-rotation coupling, {\it i.e.}, the interaction
of the graviton spin with the gravitomagnetic fields is actually a
self-interaction of the spacetime (gravitational fields). In the
present paper, we will show that this self-interaction will also
arise in a non-inertial frame of reference itself: specifically, a
rotating frame that experiences a fluctuation of its rotational
frequency ({\it i.e.}, the change in the rotational angular
frequency) will undergo a weak self-interaction. The
self-interaction of the rotating frame, which can also be called
the self-interaction of the spacetime of the rotating frame, is
just the non-inertial generalization of the interaction of the
graviton spin with the gravitomagnetic fields.
\end{abstract}
\section{Introduction}
Historically, the weak gravitational effects and the magnetic-type
gravitational fields (referred to as the gravitomagnetic fields)
have captured much attention of a number of
investigators\cite{Carmi,Bohm,Post,Horne,Mashhoon1988,Cai1991,Li,Peng,Podkletnov,Unnikrishnan,GRG}.
These effects include the Aharonov-Carmi effect\cite{Carmi}
(gravitational Aharonov-Bohm effect)\cite{Bohm}, Sagnac-type
effect\cite{Post,Horne}, spin-rotation
coupling\cite{Mashhoon1988,Cai1991}, and the effects {\it and}
phenomena arising from the interaction between gravity (including
the inertial force) {\it and}
superconductors\cite{Li,Peng,Podkletnov,Unnikrishnan}. It is now
well known that many analogies can be drawn between gravity and
electromagnetic force in some aspects, for example, in a rotating
reference frame, a moving particle is acted upon by both the
inertial centrifugal force and the Coriolis force, which are
analogous to the electric force and the Lorentz magnetic force,
respectively, in electrodynamics. For this reason, Aharonov and
Carmi proposed a geometric effect related to the vector potential
of inertial force\cite{Carmi}, and Anandan\cite{Anandan} {\it and}
Dresden {\it et al}.\cite{Dresden} (independently) proposed a
quantum-interferometry effect associated with the gravity. In
fact, these two effects are the gravitational analog to the
well-known Aharonov-Bohm effect, namely, the matter wave
propagating along a closed path in a rotating frame will acquire a
nonintegral phase factor (geometric phase factor). Thus, this
effect can also be called the gravitational Aharonov-Bohm effect.
Overhauser {\it et al}.\cite{Overhauser} and Werner{\it et
al}.\cite{Werner} (independently) have shown experimentally the
existence of the above effects by means of the neutron-gravity
interferometry experiments.

It should be noted that the gravitational Aharonov-Bohm effect
(Aharonov-Carmi effect) results from the interaction between the
momentum of a particle and the rotating frame, the rotational
frequency of which can be considered as a gravitomagnetic field:
specifically, the Coriolis force $2m\vec{v}\times\vec{\omega}$
experienced by a moving particle inside a rotating frame is just
the gravitational analog to the Lorentz magnetic force
$q\vec{v}\times\vec{B}$ acting upon a charged particle in a
magnetic field. By analogy with the interaction of the spinning
magnetic moment with the magnetic field, one may propose a
gravitational counterpart in gravity theory. Historically,
Mashhoon {\it et al.} considered such a weak gravitational effect,
{\it i.e.}, the so-called spin-rotation coupling, which means that
a spinning particle inside a rotating frame of reference will
undergo a coupling of its spin to the rotational
frequency\cite{Mashhoon1988,Cai1991,Mashhoon1,Mashhoon2,Mashhoon3}.
A remarkable feature of this interaction lies in that the
spin-rotation coupling leads to an inertial effects of the
intrinsic spin of the particle, for instance, although the
equivalence principle still holds for a spinning particle, the
universality of Galileo's law of freely falling particles is
violated, provided that the spin is polarized vertically up or
down in the non-inertial frame: specifically, a spinning particle
(with spin $\vec{s}$) in a rotating frame (with the rotating
angular velocity being $\vec{\omega}$) will undergo a force
$-\nabla(\vec{\omega}\cdot\vec{s})$. The signs of the two forces
$-\nabla(\vec{\omega}\cdot\vec{s})$ acting upon the two spinning
particles polarized vertically up and down are just opposite. For
this reason, Mashhoon concluded that the spin possesses an
inertial property and the universality of Galileo's law of freely
falling particles may therefore be violated\cite{Mashhoon2}.

The interaction between the weak gravity and the moving
superconductors is of physical
interest\cite{Li,Peng,Podkletnov,Unnikrishnan}\footnote{If
Podkletnov's experimental result (so-called weak gravitational
shielding effect)\cite{Podkletnov} is true, it can be interpreted
by using the concept of ``oscillatorily varying Meissner effect":
specifically, the gravity in the superconductor is governed by the
equation $\nabla^{2}{\bf g}+\lambda {\bf g}=0$ ($\lambda>0$),
where $\lambda$ is a parameter associated with the cosmological
constant and the self-induced mass current. In the superconducting
medium, the cosmological constant is small compared with the
contribution of the self-induced mass current, and can therefore
be ignored. The solution of this equation is ${\bf g}={\bf
g}_{0}\cos(\sqrt{\lambda}z)$, which is approximately equal to
${\bf g}_{0}\left(1-\lambda z^{2}/2\right)$. The small term $-{\bf
g}_{0}\left(\lambda z^{2}/2\right)$ may be considered a weak
gravitational shielding effect of the superconductors. }. During
the past two decades, many researchers investigated such an
interaction both theoretically and experimentally. Peng {\it et
al}. suggested a unified phenomenological theory to treat the
interaction between arbitrarily moving superconductors and
gravitational fields including the Newtonian gravity,
gravitational waves, vector transverse gravitoelectric fields, and
vector gravitomagnetic fields\cite{Peng}. In the limit of weak
field and low velocity, they considered the properties of induced
electromagnetic and gravitational fields in the interior of a
moving superconductor. It was shown that many physically
interesting effects, phenomena and properties, including the
Meissner effect, London moment, DeWitt effect, effects of
gravitational wave on a superconductor, and induced electric
fields in the interior of a freely vibrating superconductor, could
be recovered from the theoretical expressions for the induced
electromagnetic and gravitational fields in their unified
phenomenological theory\cite{Peng}. By further analysis one can
demonstrate that the weak equivalence principle is valid in
superconductivity, that Newtonian gravity and gravitational waves
will penetrate either a moving superconductor or a superconductor
at rest, and that a superconductor at rest cannot shield either
vector gravitomagnetic fields or vector transverse gravitoelectric
fields\cite{Peng}\footnote{This means that the traditional
``exponentially decay Meissner effect" is absent. However, the
so-called ``oscillatorily varying Meissner effect" exists. Also
see ``On some weak-gravitational effects" (J.Q. Shen, arXiv:
gr-qc/0305094).}. It is worth stating that Ciubotariu {\it et al}.
investigated the gravitomagnetic effects on a superconductor in
the framework of the weak stationary gravitational field and low
velocity, and obtained the similar conclusions\cite{Ciubotariu}.
Li {\it et al}. reported the investigation of the effects of a
pure superconductor on the external gravitomagnetic and magnetic
fields in a weak-gravity and low-velocity system\cite{Li}. It was
found that a small residual uniform magnetic field will pervade
the superconductor and that the external fields mutually
``induce'' additional small internal perturbation fields. Li {\it
et al}. demonstrated that their obtained results might differ from
the previous London theory and Meissner effect: the magnetic field
inside a superconductor although very small no longer vanishes,
and this nonzero internal magnetic field raises an interesting
consequence regarding the internal gravitomagnetic field, {\it
i.e.}, for a regular superconducting material, the magnetically
produced gravitomagnetic field is order of about $10^{11}$ times
the internal magnetic field\cite{Li}.

Historically, the concept of spatial gravitational forces modelled
after the electromagnetic Lorentz force has a long time and many
names associated with
it\cite{Landau,Cattaneo,Massa,Braginsky,Braginsky2,Damour}. Born
in the Newtonian context of centrifugal and Coriolis forces
introduced by a rigidly rotating coordinate system in a flat
Euclidean space, it has found a number of closely related but
distinct generalizations within the context of general relativity
and its linearized approximation. Since the frequent reference to
``gravitoelectromagnetism'' occurs in recent literature, Jantzen
{\it et al}. placed all of these notions of ``non-inertial
forces'' into a single framework which in turn might be used to
infer relationships among them\cite{Jantzen}. In addition, other
interesting study of the weak gravitational field includes the
optical-mechanical analogy in general relativity, {\it i.e.}, the
gravitational field can be represented as an optical medium with
an effective optical index of refraction\cite{Felice}. Thus, it
was shown that the orbits of both massive and massless particles
are governed by a variational principle which involves the
refractive index and which assumes the form of Fermat's principle
or of Maupertuis's principle\cite{Evans}.

In this paper, we discuss the related topics on the graviton spin
and gravitomagnetic fields. In Sec. II, we obtain the expression
for the spin of the weak gravitational field by using the
canonical procedure and Noether theorem. In Sec. III, we discuss
the connection between the rotational frequency of a non-inertial
frame and the gravitomagnetic field. In Sec. IV, we predict that
there exists a self-interaction of the rotating frame of reference
due to the fluctuation of the rotational frequency.

\section{The spin of gravitational fields}
In this paper, we deal with the weak gravitational fields only,
which is described by the linearized gravity theory. One speaks of
a linearized theory when the metric deviates only slightly from
that of flat space. A weak gravitational field (in which spacetime
is nearly flat) is defined as a manifold on which coordinates
exist, where the metric has components
$g_{\alpha\beta}=\eta_{\alpha\beta}+h_{\alpha\beta}$ with
$|h_{\alpha\beta}|\ll 1$. Such coordinates are called nearly
Lorentz coordinates, where the indices of tensors are raised and
lowered with the flat-Minkowski metric $\eta_{\mu\nu}$ and
$\eta^{\mu\nu}$ rather than with ${g}_{\mu\nu}$ and
${g}^{\mu\nu}$.

In the linearized gravity theory, there are two fundamental types
of coordinate transformations, which take one nearly Lorentz
coordinate system into another. The two coordinate transformations
are the background Lorentz transformation (the background metric
takes the form of simple diagonal $(+1,-1,-1,-1)$) and the gauge
transformation (for $h_{\mu\nu}$). For the weak gravitational
field, we can see that, under a background Lorentz transformation,
$h_{\mu\nu}$ transforms as if it were a tensor in SR all by
itself. Of course it is not a tensor. But this tells us that we
can think of a slightly curved spacetime as a flat one with a
``tensor'' $h_{\mu\nu}$ defined on it, and moreover, all physical
fields such as $R_{\alpha\beta\mu\nu}$ can be defined in terms of
$h_{\mu\nu}$, and they will look like physical fields on a flat
background spacetime\cite{Schutz}. The gauge transformation is
such that a very small change in coordinates, which is of the form
$x^{\bar{\alpha}}=x^{\alpha}+\zeta^{\alpha}(x^{\beta})$, is
generated by a vector $\zeta^{\alpha}$, the components of which
are functions of position. If we demand that the ``vector''
$\zeta^{\alpha}$ be small in the sense that the relation
$|\zeta^{\alpha},_{\beta}|\ll 1$ is satisfied, then we
have\cite{Schutz}
\begin{equation}
\Lambda^{\bar{\alpha}}_{\ \beta}=\frac{\partial
x^{\bar{\alpha}}}{\partial x^{\beta}}=\delta^{\alpha}_{\
\beta}+\zeta^{\alpha},_{\beta}, \quad \Lambda^{{\alpha}}_{\
\bar{\beta}}=\frac{\partial x^{{\alpha}}}{\partial
x^{\bar{\beta}}}=\delta^{\alpha}_{\
\beta}-\zeta^{\alpha},_{\beta}+{\mathcal
O}\left(|\zeta^{\alpha},_{\beta}|^{2}\right),  \label{eeqeq21}
\end{equation}
By using the background Lorentz transformation
$g_{\bar{\alpha}\bar{\beta}}=\Lambda^{\alpha}_{\
\bar{\alpha}}\Lambda^{\beta}_{\ \bar{\beta}}g_{\alpha\beta}$,
keeping in mind Eq.(\ref{eeqeq21}), one can easily verify that, to
the first order in small quantities,
\begin{equation}
g_{\bar{\alpha}\bar{\beta}}=\eta_{\alpha\beta}+h_{\alpha\beta}-\zeta_{\alpha,\beta}-\zeta_{\beta,\alpha},
\end{equation}
where we define $\zeta_{\alpha}=\eta_{\alpha\beta}\zeta^{\beta}$.
This means that the effect of the coordinate change is to
re-define $h_{\alpha\beta}$:
\begin{equation}
h_{\alpha\beta}\rightarrow
h_{\alpha\beta}-\zeta_{\alpha,\beta}-\zeta_{\beta,\alpha},
\label{gauge transformation}
\end{equation}
which can be considered as a gauge transformation. Such a gauge
transformation enables us to choose an appropriate gauge condition
so as to simplify the expressions for the physical quantities
(including the canonical momentum, spin) of gravitational fields.

In the following, we will obtain the expression for the spin of
gravitational fields by making use of the Noether theorem and
canonical procedure with the linearized gravity theory.

In the linearized gravity theory, to the first order in
$h_{\mu\nu}$, the Riemann tensor is written
\begin{equation}
R_{\alpha\beta\mu\nu}=\frac{1}{2}\left(h_{\alpha\nu,\beta\mu}+h_{\beta\mu,\alpha\nu}-h_{\alpha\mu,\beta\nu}-h_{\beta\nu,\alpha\mu}\right)+{\mathcal
O}(h^{2}),
\end{equation}
and consequently the curvature scalar is of the form
\begin{equation}
R=g^{\alpha\mu}g^{\beta\nu}R_{\alpha\beta\mu\nu}=2h^{\beta\nu}\left(h^{\mu}_{\
\ \nu,\beta\mu}-\frac{1}{2}h_{\beta\nu,\mu}^{\ \ \ \ \
\mu}-\frac{1}{2}h^{\mu}_{\ \ \mu,\beta\nu}\right)+{\mathcal
O}(h^{3}).
\end{equation}
Thus, the action of the gravitational field may be written in the
form
\begin{equation}
I=\int_{\Omega}{\rm d}\Omega\sqrt{-g}R\simeq \int_{\Omega}{\rm
d}\Omega\left[-2\left(h^{\beta\nu}_{\ \ \ ,\mu}h^{\mu}_{\ \
\nu,\beta}-\frac{1}{2}h^{\beta\nu,\mu}h_{\beta\nu,\mu}-\frac{1}{2}h^{\beta\nu}_{\
\ \ ,\nu}h^{\mu}_{\ \ \mu,\beta}\right)\right]+{\rm S.T.},
\end{equation}
where ${\rm S.T.}$ denotes the surface term. It follows that, in
the nearly Lorentz coordinate system, the Lagrangian density of
the gravitational field takes the form
\begin{equation}
{\mathcal
L}=-2\left(h^{\beta\nu,\mu}h_{\mu\nu,\beta}-\frac{1}{2}h^{\beta\nu,\mu}h_{\beta\nu,\mu}-\frac{1}{2}h^{\beta\nu}_{\
\ \ ,\nu}h^{\mu}_{\ \ \mu,\beta}\right).   \label{lagrangian}
\end{equation}
According to the canonical procedure, the canonical momentum of
the linearized gravitational field is
\begin{equation}
\pi^{\mu\nu}=\frac{\partial {\mathcal L}}{\partial
\dot{h}_{\mu\nu}}=-4\left(h^{0\nu,\mu}-\frac{1}{2}h^{\mu\nu,0}\right)+\eta^{\mu\nu}h^{0\lambda}_{\
\ \ ,\lambda}+\eta^{\nu 0}h^{,\mu},    \label{pi}
\end{equation}
where dot denotes the derivative with respect to time, and
$h=\eta^{\mu\nu}h_{\mu\nu}$. In accordance with the Noether
theorem, the spin of a field that is characterized by the
Lagrangian density (\ref{lagrangian}) takes the form
\begin{equation}
S^{\theta\tau}=\int {\rm
d}^{3}x\left(\pi^{\mu\nu}\sum^{\theta\tau}_{\nu\eta}h^{\eta}_{\ \
\mu}\right)    \quad {\rm with}    \quad
\sum^{\theta\tau}_{\nu\eta}=\delta^{\theta}_{\nu}\delta^{\tau}_{\eta}-\delta^{\theta}_{\eta}\delta^{\tau}_{\nu}.
\end{equation}
Now we calculate the integrand $s^{\theta\tau}$ of the above
integral. With the help of (\ref{pi}), one can arrive at
\begin{equation}
s^{\theta\tau}=\pi^{\mu\nu}\sum^{\theta\tau}_{\nu\eta}h^{\eta}_{\
\
\mu}=\left[-4\left(h^{0\nu,\mu}-\frac{1}{2}h^{\mu\nu,0}\right)+\eta^{\mu\nu}h^{0\lambda}_{\
\ \ ,\lambda}+\eta^{\nu
0}h^{,\mu}\right]\left(\delta^{\theta}_{\nu}\delta^{\tau}_{\eta}-\delta^{\theta}_{\eta}\delta^{\tau}_{\nu}\right)h^{\eta}_{\
\ \mu}.
\end{equation}
Note that here the indices $\theta,\tau$ take values from $1$ to
$3$. Further calculation yields
\begin{equation}
s^{\theta\tau}=-4\left(h^{0\theta,\mu}h^{\tau}_{\ \
\mu}-h^{0\tau,\mu}h^{\theta}_{\ \
\mu}\right)+2\left(h^{\mu\theta,0}h^{\tau}_{\ \
\mu}-h^{\mu\tau,0}h^{\theta}_{\ \ \mu}\right)
+\left(h^{0\lambda}_{\ \ \ ,\lambda}h^{\theta\tau}-h^{0\lambda}_{\
\ \ ,\lambda}h^{\tau\theta}\right)+h^{,\mu}\left(\eta^{\theta
0}h^{\tau}_{\ \ \mu}-\eta^{\tau 0}h^{\theta}_{\ \ \mu}\right).
\label{s}
\end{equation}
It is apparently seen that the third term on the right-handed side
of the expression (\ref{s}) vanishes, {\it i.e.}, $h^{0\lambda}_{\
\ \ ,\lambda}h^{\theta\tau}-h^{0\lambda}_{\ \ \
,\lambda}h^{\tau\theta}=0$, since $h^{\theta\tau}$ is symmetric in
indices. Additionally, here we will choose a gauge condition in
which the relation $h_{00}=h_{11}=h_{22}=h_{33}=0$ is satisfied.
According to the gauge transformation (\ref{gauge
transformation}), this condition (containing four relations) can
be easily realized only by introducing an appropriate ``gauge
vector'' (vector parameters) $\zeta_{\alpha}$ ($\alpha$ runs from
0 to 3)\footnote{The introduction of $\zeta_{\alpha}$ means that
the four gauge conditions are introduced and can fix the four
gravitational field components $h_{\mu\nu}$, say, $h_{00}, h_{11},
h_{22}, h_{33}$. According to the gauge transformation (\ref{gauge
transformation}), the four gauge conditions are $h_{ii}\rightarrow
h_{ii}-2\zeta_{i,i}=0$ with $i=0,1,2,3$. Thus, in this gauge
condition, all the diagonal components $h_{ii}=0$ ($i=0,1,2,3$).}.
This, therefore, means that $h=\eta^{\mu\nu}h_{\mu\nu}=0$ and, in
consequence, $h^{,\mu}=0$. Thus, the fourth term on the
right-handed side of the expression (\ref{s}) is also vanishing.
Hence, it follows from (\ref{s}) that the density of the spin of
the weak gravitational field is rewritten as
\begin{equation}
s^{\theta\tau}=-4\left(h^{0\theta,\mu}h^{\tau}_{\ \
\mu}-h^{0\tau,\mu}h^{\theta}_{\ \
\mu}\right)+2\left(h^{\mu\theta,0}h^{\tau}_{\ \
\mu}-h^{\mu\tau,0}h^{\theta}_{\ \ \mu}\right).
\end{equation}

In what follows we will let the derivatives of the ``gauge
vector'' (vector parameters) $\zeta_{\alpha}$ agree with the
following three conditions $h_{ij}\rightarrow
h_{ij}-\zeta_{i,j}-\zeta_{j,i}\rightarrow 0$ ($i,j=1,2,3$), which
means that the off-diagonal spacial-part field components $h_{12},
h_{23}, h_{31}$ approach $0$, and the only retained components are
the gravitomagnetic vector potentials $h_{0i}$ ($i=1,2,3$). It
should be noted that this gauge condition can be readily realized
in the linearized gravity theory, for example, for the rotating
gravitating body, the gravitomagnetic components $h_{0i}$
($i=1,2,3$) are in general much greater than the other
off-diagonal field components $h_{ij}$ ($i,j=1,2,3$), that is,
there exists a coordinate system in which the off-diagonal field
components $h_{ij}$ ($i,j=1,2,3$) vanishes (or nearly vanishes).
Hence, in this gauge condition, the spin of gravitational field
can be simplified as the following form $
s^{\theta\tau}=-2\left(h^{0\theta,0}h^{\tau
0}-h^{0\tau,0}h^{\theta 0}\right)$. If the gravitomagnetic vector
potentials are represented by a three-dimensional ``vector''
$g^{\theta}$, {\it i.e.}, $h^{0\theta}=g^{\theta}$, then the above
expression for $s^{\theta\tau}$ can be rewritten as
\begin{equation}
s^{\theta\tau}=-2\left(\dot{g}^{\theta}g^{\tau}-\dot{g}^{\tau}g^{\theta}\right),
\label{2factor}
\end{equation}
where dot denotes the time derivative. Note that in the
electrodynamics, the spin of the electromagnetic field is
expressed by $s^{\theta\tau}_{\rm
e.m.}=-\left(\dot{A}^{\theta}A^{\tau}-\dot{A}^{\tau}A^{\theta}\right)$\cite{Bjorken}.
It follows from the comparison of the expression for the spin of
electromagnetic field with (\ref{2factor}) that the factor $2$ in
(\ref{2factor}) may imply that the spin of a weak gravitational
field is just two times that of the electromagnetic field.

We have studied a purely gravitational generalization of
spin-rotation couplings, the Lagrangian density of which
is\cite{PRD}\footnote{For the detailed derivation, see, for
example, ``The interaction between graviton spin and
gravitomagnetic fields" (J.Q. Shen, arXiv: gr-qc/0312116).}
\begin{equation}
{\mathcal L}_{\rm
s-g}=\frac{1}{2}K^{3}\left(g_{i}\dot{g}_{j}-g_{j}\dot{g}_{i}\right)\left(\partial_{i}g_{j}-\partial_{j}g_{i}\right),
\label{Lag}
\end{equation}
where $i,j=1, 2, 3$, and the coefficient $K$ and the
gravitomagnetic vector potential $g_{i}$ are so defined that the
relation $\sqrt{-g}g_{0i}\simeq Kg_{i}$ is satisfied. Since the
expression $\left(g_{i}\dot{g}_{j}-g_{j}\dot{g}_{i}\right)$ in
(\ref{Lag}) can be thought of as a term associated with the spin
of gravitational field, and $\partial_{i}g_{j}-\partial_{j}g_{i}$
is an expression for the gravitomagnetic field, the nature of such
a purely gravitational generalization is an interaction between
the gravitational spinning moment (gravitomagnetic moment) and the
gravitomagnetic fields. In the next section, we will show that for
the spacetime in the rotating frame of reference, the expression
$\partial_{i}g_{j}-\partial_{j}g_{i}$ contains the angular
velocity of the rotating frame, {\it i.e.}, the rotational
frequency is only a piece of the gravitomagnetic field. Hence, it
is shown that the interaction in (\ref{Lag}) is truly a purely
gravitational generalization of Mashhoon's spin-rotation
couplings.

\section{Mashhoon's spin-rotation coupling and gravitomagnetic fields}

Mashhoon showed that a spinning particle in a rotating frame of
reference will experience an interaction between its spin and the
angular frequency of the rotating frame. This interaction is
referred to as the spin-rotation
coupling\cite{Mashhoon1988,Cai1991,Mashhoon1,Mashhoon2,Mashhoon3}.
It is clearly seen that this coupling is in fact an inertial
effect of spinning particles.

Further analysis may show that here the spin of the spinning
particle can be regarded as the gravitomagnetic moment, and the
angular velocity of the rotating frame can be thought of as a
gravitomagnetic field\footnote{These viewpoints are adopted from
Ref.\cite{Shen}.}, since it can be verified that, in a rotating
frame of reference, the spin-rotation coupling is involved in the
interaction of the gravitomagnetic moment and the gravitomagnetic
field.

In the following, by taking account of the coordinate
transformation from the rest frame to the rotating frame, we will
show that, for a moving particle, the rotating angular frequency
of a rotating frame of reference can truly be viewed as a
gravitomagnetic field. In order to deal with this problem, we
consider the Kerr metric of the exterior gravitational field of
the rotating spherically symmetric body, which is of the form

\begin{eqnarray}
{\rm d}s^{2} &=&\left[1-\frac{2GMr}{c^{2}(r^{2}+a^{2}\cos
^{2}\theta )}\right]c^{2}{\rm d}t^{2}-\frac{r^{2}+a^{2}\cos
^{2}\theta }{r^{2}+a^{2}-\frac{2GMr}{c^{2}}}{\rm
d}r^{2}-(r^{2}+a^{2}\cos ^{2}\theta ){\rm d}\theta ^{2}
\nonumber \\
& & -\sin ^{2}\theta \left(\frac{2a^{2}\sin ^{2}\theta
}{r^{2}+a^{2}\cos^{2}\theta
}\frac{GMr}{c^{2}}+r^{2}+a^{2}\right){\rm d}\varphi
^{2}+\frac{2a\sin ^{2}\theta }{r^{2}+a^{2}\cos ^{2}\theta
}\frac{GMr}{c}{\rm d}t{\rm d}\varphi , \label{eq2}
\end{eqnarray}
where $r,\theta ,\varphi $ are the displacements of spherical
coordinate. Here $a$ is so defined that $ac$ is the angular
momentum of unit mass of the rotating gravitating body, and $M$
denotes the mass of this gravitating body. Note that the spacetime
coordinate of Kerr metric (\ref{eq2}) is in the rest (fixed)
reference frame. We can transform the above Kerr metric into the
form in the rotating reference frame. Because of the smallness of
the Earth$^{,}$s rotating velocity, one can apply the following
Galileo transformation to the coordinates of a test particle
moving radially in the rotating frame

\begin{equation}
{\rm d}r^{^{\prime }}+v{\rm d}t^{^{\prime }}={\rm d}r,  \quad {\rm
d}\theta ^{^{\prime }}={\rm d}\theta ,  \quad   {\rm d}\varphi
^{^{\prime }}={\rm d}\varphi +\omega {\rm d}t,  \quad
 {\rm d}t^{^{\prime }}={\rm d}t \label{eq3}
\end{equation}
with $v$ being the radial velocity of the test particle relative
to the rotating reference frame, $(r^{^{\prime }},\theta
^{^{\prime }},\varphi ^{^{\prime }},t^{^{\prime }})$ and
$(r,\theta ,\varphi ,t)$ the spacetime coordinates of the rotating
frame and fixed frame, respectively. $\omega $ denotes the
rotational frequency of the rotating frame with respect to the
fixed reference frame. For the simplicity of calculation, the
radial velocity $v$
is taken to be much less than $\omega r$. The substitution of Eq.(\ref{eq3}%
) into Eq.(\ref{eq2}) yields

\begin{eqnarray}
{\rm d}s^{2} &=&\left[1-\frac{2GMr}{c^{2}(r^{2}+a^{2}\cos
^{2}\theta )}-\frac{ (r^{2}+a^{2}\cos ^{2}\theta
)}{r^{2}+a^{2}-\frac{2GMr}{c^{2}}}\frac{v^{2}}{ c^{2}}-\sin
^{2}\theta \left(r^{2}+a^{2}+\frac{2a^{2}\sin ^{2}\theta }{
r^{2}+a^{2}\cos ^{2}\theta }\frac{GMr}{c^{2}}
-\frac{2a}{r^{2}+a^{2}\cos ^{2}\theta }\frac{GMr}{\omega
c}\right)\frac{\omega
^{2}}{c^{2}}\right]     \nonumber \\
&& \times c^{2}{\rm d}t^{^{\prime }2}-\frac{r^{2}+a^{2}\cos
^{2}\theta }{ r^{2}+a^{2}-\frac{2GMr}{c^{2}}}{\rm d}r^{^{\prime
}2}-(r^{2}+a^{2}\cos ^{2}\theta ){\rm d}\theta ^{^{\prime }2}-\sin
^{2}\theta \left( \frac{2a^{2}\sin ^{2}\theta }{r^{2}+a^{2}\cos
^{2}\theta }\frac{GMr}{c^{2}}+r^{2}+a^{2}\right){\rm d}\varphi
^{^{\prime }2}
\nonumber \\
& & -\frac{2(r^{2}+a^{2}\cos ^{2}\theta )}{
r^{2}+a^{2}-\frac{2GMr}{c^{2}}}\frac{v}{c}{\rm d}r^{^{\prime
}}c{\rm d}t^{^{\prime }}
+\left[\frac{2a\sin ^{2}\theta }{r^{2}+a^{2}\cos ^{2}\theta }\frac{GMr}{c}%
+2\sin ^{2}\theta \left(r^{2}+a^{2}+\frac{2a^{2}\sin ^{2}\theta
}{r^{2}+a^{2}\cos ^{2}\theta }\frac{GMr}{c^{2}}\right)\omega
\right]{\rm d}t^{^{\prime }}{\rm d}\varphi ^{^{\prime }},
\label{eq4}
\end{eqnarray}
where the term $\frac{\omega ^{2}r^{2}}{c^{2}}\sin ^{2}\theta $ in
$g_{tt}$ results in the inertial centrifugal force written as
$\vec{F}=m\vec{\omega}\times (\vec{\omega}\times \vec{r})$. Ignoring the
small terms associated with the relation $\frac{a^{2}%
}{r^{2}}\ll 1$ in $g_{t\varphi ^{^{\prime }}},$ one can arrive at

\begin{eqnarray}
g_{t\varphi ^{^{\prime }}}{\rm d}\varphi ^{^{\prime }}{\rm
d}t^{^{\prime }} &=&\left(\frac{ 2aGMr\sin ^{2}\theta
}{cr^{2}}+2\omega r^{2}\sin ^{2}\theta \right){\rm d}t^{^{\prime
}}{\rm d}\varphi ^{^{\prime }}  \nonumber \\
&=&\left(\frac{2aGM\sin \theta }{cr^{2}}+2\omega r\sin \theta
\right){\rm d}t^{^{\prime }}r\sin \theta {\rm d}\varphi ^{^{\prime
}}. \label{eq5}
\end{eqnarray}
Thus, the gravitomagnetic potentials can be written as
\begin{equation}
g_{\varphi }=\frac{2aGM\sin \theta }{cr^{2}}+2\omega r\sin \theta
,\quad g_{r}=-2v,  \quad   g_{\theta }=0.  \label{eq6}
\end{equation}
It follows that the first term, $\frac{2aGM\sin \theta }{cr^{2}}$,
on the right-handed side of (\ref{eq6}) is exactly analogous to
the magnetic potential $\frac{\mu _{0} }{4\pi
}\frac{ea}{r^{2}}\sin \theta $ produced by a rotating charged
spherical shell in the electrodynamics. So, the first term
$\frac{2aGM\sin \theta }{cr^{2}}$ of $g_{\varphi }$ shows the
existence of a gravitomagnetic field of the rotating gravitating
body, the exterior gravitomagnetic strength of which
is\cite{Ahmedov}
\begin{equation}
\vec{B}_{\rm
g}=\frac{2G}{c}\left[\frac{\vec{a}}{r^{3}}-\frac{3\left(\vec{a}\cdot
\vec{ r}\right)\vec{r}}{r^{5}}\right].
\end{equation}

In accordance with the equation of geodesic line of a particle in
the post-Newtonian approximation, the gravitomagnetic strength can
be defined by $-\frac{1}{2}\nabla \times \vec{g}$ with
$\vec{g}=(g_{01},g_{02},g_{03})$ as assumed above. If we set \
$\beta _{\varphi }=2\omega r\sin \theta ,\beta _{r}=-2v,\beta
_{\theta }=0$, then the gravitomagnetic strength that arises in
the rotating reference frames is given as follows

\begin{equation}
-\frac{1}{2}\nabla \times \vec{\beta}=-2\omega \cos \theta
e_{r}+2\omega \sin \theta e_{\theta }  \label{eq7}
\end{equation}
with $e_{r},e_{\theta }$ being the unit vector. It follows from Eq.(\ref{eq7}%
) that such a gravitomagnetic strength is related to the rotation
of non-inertial frame and independent of the Newtonian
gravitational constant $ G$. From the point of view of Newtonian
mechanics, it is the inertial force field in essence rather than
the field produced by mass current. Since we have assumed that the
velocity of a test particle is parallel to $e_{r}$, {\it i.e.},
$\vec{v}=ve_{r},$ the gravitational Lorentz force acting on the
test particle in the gravitomagnetic field is thus given by
\begin{equation}
\vec{F}=m\vec{v}\times \left(-\frac{1}{2}\nabla \times
\vec{\beta}\right)=2v\omega \sin \theta e_{\varphi
}=2m\vec{v}\times \vec{\omega}.  \label{eq8}
\end{equation}
We conclude from Eq.(\ref{eq8}) that the gravitational Lorentz
force in a rotating reference frame is just the familiar Coriolis
force, and that the rotational frequency $\vec{\omega}$ can be
regarded as a gravitomagnetic field strength (or a piece of the
gravitomagnetic field strength).

Since the angular frequency of the rotating frame is just a piece
of the gravitomagnetic field, Mashhoon's spin-rotation coupling is
actually the interaction of the gravitomagnetic moment with the
gravitomagnetic field. In the literature, to the best of our
knowledge, the spin-rotation coupling of photon, electron and
neutron has been taken into account\cite{Mashhoon2,Shen,He}.
However, the gravitational coupling of graviton spin to the
gravitomagnetic fields, which may also be of physical interest,
has so far never yet been considered. We think it is necessary to
extend Mashhoon's spin-rotation coupling to a purely gravitational
case, where the graviton spin will be coupled to the
gravitomagnetic fields.

\section{The self-interaction of the non-inertial frame of reference}

For the present, Mashhoon's spin-rotation coupling could be tested
only in microwave experiments\cite{Mashhoon2}, since it is
relatively weak due to the smallness of the rotational frequencies
of various rotating frames on the Earth. We argue that besides the
microwave experiments, there is another new method to test for the
spin-rotation coupling, {\it i.e.}, the rotational motion of
C$_{60}$ molecules may provide us with an ingenious way to detect
this weakly gravitational (gravitomagnetic) effect\cite{He}. It
follows that since in the high-temperature phase (orientationally
disordered phase) of C$_{60}$ solid, the rotational frequency,
$\omega $, of C$_{60}$ molecules may be about $10^{11}$
rad/s\cite{He}, the precessional frequency $\Omega $ is therefore
compared to $\omega$, {\it i.e.},
\begin{equation}
\Omega \simeq \frac{|\vec {M}| }{I\omega }
\end{equation}
ranges from $10^{10}$ to $10^{12}$ rad/s. Here, $\vec {M}$ and $I$
denote the moment of noncentral intermolecular force and the
C$_{60}$ moment of inertia, respectively. Because the rotational
angular velocity is much greater than that of any rotating bodies
on the Earth, the C$_{60}$ molecule is an ideal non-inertial frame
of reference, where the effects for the valency electrons in the
C$_{60}$ molecule resulting from the electron spin-rotation
coupling may be easily observed experimentally.

In addition to the above electron spin-rotation coupling in the
rotating C$_{60}$ molecule, there may exist another novel
interaction, which is related closely to the inertial property of
the rotating frame of reference itself. The linearized gravity
theory indicates that the interaction of the graviton spin with
the gravitomagnetic fields (a purely gravitational generalization
of Mashhoon's spin-rotation coupling) is a self-interaction of the
spacetime (gravitational fields). Here, we will show that this
self-interaction will also arise in a non-inertial frame of
reference itself, namely, a rotating frame, the rotating frequency
of which fluctuates, will undergo a weak self-interaction
described by (\ref{Lag}). It is believed that such a
self-interaction of the rotating frame is just the non-inertial
generalization of the interaction of the graviton spin with the
gravitomagnetic fields ({\it i.e.}, the self-interaction of the
spacetime). According to Sec. 2, in a rotating frame, the
$\varphi$-component of the gravitomagnetic vector potential is
$g_{\varphi }=\frac{2aGM\sin \theta }{cr^{2}}+2\omega r\sin
\theta$. Generally speaking, in the weak gravitational field, the
contribution of the term $2\omega r\sin \theta$ is much greater
than that of $\frac{2aGM\sin \theta }{cr^{2}}$. So, the
$\varphi$-component of the gravitomagnetic vector potential
(measured by the observer fixed in the C$_{60}$ rotating frame) is
$ g_{\varphi }\simeq 2\omega r\sin \theta$. It follows that if
either the direction or the magnitude of the angular velocity of
the rotating frame changes, then the time derivative of the
gravitomagnetic vector potentials is nonvanishing, {\it i.e.},
\begin{equation}
|\dot{g}_{\varphi }|\simeq |2\dot{\omega} r\sin \theta|\neq 0,
\end{equation}
and according to (\ref{Lag}), this rotating frame will be
subjected to a self-interaction, the nature of which is just a
nonlinear interaction of the spacetime of the non-inertial frame
of reference itself. It is apparent that such a self-interaction
of the non-inertial frame of reference is one of the predictions
of both Einstein's field equation and the principle of
equivalence, while in the Newtonian mechanics, there exists no
such a self-interaction.

In what follows, we will consider the change in the angular
frequencies of rotating C$_{60}$ molecules acted upon by a
noncentral intermolecular force that causes both the precession
and nutation of molecular rotation of C$_{60}$. Roughly speaking,
the moment of noncentral intermolecular force, $\vec{M}$, is the
product of van der Waals force (referring to the noncentral part)
and the distance between molecules, the order of magnitude
$|\vec{M}|$ of which is $10^{-22}\sim 10^{-20}$ N$\cdot$m. The
equation of the rotational motion of C$_{60}$ molecule coupled to
its neighbors is $\frac{\rm d}{{\rm d}t}\vec{L}=\vec{M}$ with the
angular momentum $\vec{L}=I\vec{\omega}$, where the C$_{60}$
moment of inertia is $I\simeq 1.0\times 10^{-43}$ Kg$\cdot{\rm
m}^{2}$. Thus the rate of change of the C$_{60}$ angular velocity
$\vec{\omega}$ is\footnote{Such a large fluctuation of angular
velocity (gravitomagnetic field) of frame of reference will
generate a kind of space-time ripples, which can be called the
non-inertial gravitational wave. Detecting and investigating the
gravitational wave is one of the leading areas in both
astrophysics and cosmology. Because of the weakness of the
gravitational radiation, gravitational wave has not been detected
up to now even by means of both resonant-mass detectors and
laser-interferometric detectors. Here, however, the density of the
non-inertial gravitational wave is no longer restricted to the
smallness of the gravitational constant. For the detailed
discussion of the non-inertial gravitational wave, see ``On some
weak-gravitational effects" (J.Q. Shen, arXiv: gr-qc/0305094).}
\begin{equation} \frac{\rm
d}{{\rm d}t}\vec{\omega}=\frac{1}{I}\vec{M}=\left(10^{21}\sim
10^{23}\right){\rm s}^{-2}.
\end{equation}
So, the linear acceleration of valency electrons on the C$_{60}$
molecular surface due to the above fluctuation of $\vec{\omega}$
is about $ 10^{12}\sim 10^{14}$ m$/{\rm s}^{2}$, which is the same
order of magnitude of the inertial centrifugal
acceleration\footnote{In the high-temperature phase
(orientationally disordered phase), the rotational frequency,
$\omega $, of C$_{60}$ molecules is about $10^{11}$ rad/s. The
C$_{60}$ radius is $a=3.55\AA$. So, the inertial centrifugal
acceleration, $\omega^{2}a$, experienced by the valency electrons
in C$_{60}$ molecules, is about $3\times 10^{12}$ m$/{\rm
s}^{2}$.} ($\sim 3\times 10^{12}$ m$/{\rm s}^{2}$) of the valency
electrons of C$_{60}$ due to the rotational motion of the
C$_{60}$.

It is well known that the Aharonov-Carmi effect\cite{Carmi},
neutron Sagnac effect\cite{Post,Horne}, spin-rotation
coupling\cite{Mashhoon1988,Cai1991} are the inertial effects of
the rotating frame. Most of these inertial effects may have
influence on the frequency (energy) shift in atoms, molecules and
light wave. For example, in the optical ``Foucault pendulum'' (the
Michelson-Gale light interferometer), the rotation of Earth could
yield a Sagnac shift of the light waves\cite{Post}; the energy of
the valency electrons of C$_{60}$ is shifted due to the
interaction of the electron spin with the rotation of
C$_{60}$\cite{He}. We think that since both the fluctuation in the
angular velocity of C$_{60}$ molecules and the change in the
gravitomagnetic vector potential (measured by the observer fixed
in the C$_{60}$ rotating frame) is very great, the above-mentioned
self-interaction of C$_{60}$ rotating frame may deserve
investigation for the treatment of the photoelectron spectroscopy,
noncentral intermolecular potential and molecular rotational
dynamics of C$_{60}$ solid\footnote{Additionally, maybe such a
self-interaction of the non-inertial frame of reference is also
worth considering inside the non-inertial frame itself if the
change in the angular velocity of the binary pulsar (which can
generate the gravitational radiation) and the rotating star due to
the gravitational collapse is rather great.}.
\\ \\

The consideration of the gravitomagnetic fields, the spin of the
weak gravitational field and even the gravitomagnetic dynamo
theory can lead to the appearance of many novel gravitomagnetic
effects. We hope the above-mentioned nonlinear interaction of the
rotating frame of reference itself and its potential effect on the
photoelectron spectroscopy of rotating C$_{60}$ molecules would be
studied experimentally in the near future.

\end{document}